\documentclass[3p]{elsarticle}
\usepackage{aasmacros}
\usepackage{amsmath, amssymb}
\usepackage{multicol}
\usepackage{xcolor}
\usepackage{xspace}
\usepackage{enumerate}
\usepackage{tablefootnote}
\usepackage{threeparttable}
\usepackage{booktabs}
\usepackage[english]{babel}
\usepackage[section]{placeins}
\usepackage[pdfpagelabels, plainpages=false, linktocpage=true]{hyperref}
\hypersetup{
   colorlinks   = true, %Color links instead of ugly boxes
   urlcolor     = blue, %Color for external hyperlinks
   linkcolor    = MidnightBlue, %Color of internal links
   citecolor    = red %Color of citations
}

\journal{PoS: 38th ICRC 2023}

\begin{document}

\title{Efficient Modeling of Heavy Cosmic Rays Propagation in Evolving Astrophysical Environments}

\author[1,2]{L.~Merten\corref{cor1}}%0000-0003-1332-9895
\ead{lukas.merten@rub.de}
\author[3]{P.~Da Vela}
\ead{paolo.da-vela@uibk.ac.at}
\author[3]{A.~Reimer}%0000-0001-8604-7077
\ead{anita.reimer@uibk.ac.at}
\author[3]{M.~Boughelilba}%0000-0003-1046-1647
\ead{margot.boughelilba@uibk.ac.at}
\author[4]{J.~P.~Lundquist}
\ead{jplundquist@gmail.com}
\author[4]{S.~Vorobiov}%0000-0001-8679-3424
\ead{serguei.vorobiov@gmail.com}
\author[1,2,5]{J.~Becker Tjus}
\ead{julia.tjus@rub.de}

\address[1]{Theoretical Physics IV, Plasma Astroparticle Physics, Faculty for Physics and Astronomy, Ruhr University Bochum, 44780 Bochum, Germany}
\address[2]{Ruhr Astroparticle and Plasma Physics Center (RAPP Center), Germany}
\address[3]{Universit\"at Innsbruck, Institut f\"ur Astro- und Teilchenphysik, Technikerstraße 25, 6020 Innsbruck, Austria}
\address[4]{Center for Astrophysics and Cosmology (CAC), University of Nova Gorica, Vipavska 13, SI-5000 Nova Gorica, Slovenia}
\address[5]{Department of Space, Earth and Environment, Chalmers University of Technology, 412 96 Gothenburg, Sweden}

%corresponding author
\cortext[cor1]{Corresponding author}

\begin{abstract}
We present a new energy transport code that models the time dependent and non-linear evolution of spectra of cosmic-ray nuclei, their secondaries, and photon target fields. The software can inject an arbitrary chemical composition including heavy elements up to iron nuclei. Energy losses and secondary production due to interactions of cosmic ray nuclei, secondary mesons, leptons, or gamma-rays with a target photon field are available for all relevant processes, e.g., photo-meson production, photo disintegration, synchrotron radiation, Inverse Compton scattering, and more. The resulting x-ray fluxes can be fed back into the simulation chain to correct the initial photon targets, resulting in a non-linear treatment of the energy transport. The modular structure of the code facilitates simple extension of interaction or target field models.

We will show how the software can be used to improve predictions of observables in various astrophysical sources such as jetted active galactic nuclei (AGN). Since the software can model the propagation of heavy ultrahigh-energy cosmic rays inside the source it can precisely predict the chemical composition at the source. This will also refine predictions of neutrino emissions --– they strongly depend on the chemical composition. This helps in the future to optimize the selection and analyses of data from the IceCube neutrino observatory with the aim to enhance the sensitivity of IceCube and reduce the number of trial factors.
\end{abstract}

%\begin{keyword}
%    acceleration of particles \sep radiation mechanisms: nonthermal \sep galaxies: jets \sep %galaxies: active \sep cosmic rays %from THE ASTROPHYSICAL JOURNAL SUBJECT HEADINGS
%\end{keyword}

\maketitle

\section{Introduction}
\label{sec:intro}

%Multi-messenger / multi wavelength era where time dependent source modelling becomes key to learn more about the origin of cosmic rays.

%Source often well resolved in time but so well in space --> 1D but time dependent

%Non-linear feedback plays a crucial role, when the CR should be considered too.

% No codes exists that can do all of that --> we work on an extension of CR-ENTREES which predecessors were successfully applied to SPB models.

We are at the beginning of the multi-messenger astrophsysics era. More and more astrophysical objects are observed in different wavebands, where the quasi simultaneously observation of M87 \cite{M87} is the outstanding example. For some sources we do have evidence for observations across different messengers, e.g., TXS~0506+056 \cite{TXS_IceCube}. To understand these precise observations an improved modelling of transport and radiation processes is required, that can resemble neutral messengers such as neutrinos and gamma rays. Having in mind, that the origin of ultra-high energy cosmic rays (UHECRs) is still unclear and that the composition of the cosmic-ray spectrum is one of the key observables at the highest energies, such models should also include a sophisticated description of transport and loss processes of heavy nuclei in the source. Many of the promising source candidates such as jetted active galactic nuclei (AGN) form environments, where especially the photon target fields are varying in time and are influenced by the CR population itself. In contrast to the good time resolution, the emission regions are often not well spatially resolved, which allows for a simplified one dimensional model.

In this work we describe the concept of a cosmic-ray energy transport code, that allows for non-linear, time evolving modelling of CRs up to iron including all charged and neutral secondary particles. 

\section{Theoretical Model --- Energy Transport Equation}
\label{sec:model}

As discussed in the introduction a full three dimensional model of the cosmic-ray transport and interactions is often not necessary: either because observations are limited by poor resolution of instruments or information on arrival directions is not needed, e.g., for the analysis of energy spectrum and composition of ultra-high energy cosmic rays (UHECRs). In these cases a one dimensional energy transport equation is a well motivated description of the relevant processes. It can be formulated as:
\begin{align}
    \frac{\partial f^\alpha(E, t)}{\partial t} = -\frac{\partial}{\partial E}\left(\frac{\mathrm{d} E}{\mathrm{d} t} f^\alpha \right) - \Gamma^\alpha f^\alpha + \sum_\beta Q^{\beta \alpha} f^\beta + S^\alpha \quad ,\label{eq:energytransport}
\end{align}
where $f^\alpha$ is the spectral number density of species $\alpha$, which is influenced by continues losses ($\mathrm{d}E/\mathrm{d}t$), catastrophic loss $\Gamma^\alpha$, re-injection from the same or different species $Q^{\beta \alpha}$, and sources and potential sinks $S^\alpha$. Here, the usual spatial transport terms describing diffusion and advection of particles are absorbed into an potentially energy dependent escape time $\tau_\mathrm{esc}$, which is part of $\Gamma^\alpha$. This escape time depends on the assumed transport model and the geometry of the emission region. All loss and gain terms are in general energy dependent and potentially depend on the photon target background. The photon target field is assumed to be isotropic in the emission zone frame and is discretized on a logarithmic equally space grid with 161 bins and $E_\mathrm{ph}=(10^{-19} - 10^{-3})\,\mathrm{GeV}$.

Equation \ref{eq:energytransport} can be discretized in energy $n_i=\int_{E_i^-}^{E_i^+} f(E)\,\mathrm{d}E$, where $n_i$ is the number density of a species at bin center $E_i$ derived by integrating the flux from the lower bin edge $E_i^- = E_i\times 10^{-\delta/2}$ to the upper bin edge  $E_i^+ = E_i\times 10^{+\delta/2}$. The energy spectrum is then represented by a vector of length $N_\mathrm{bin}$. For this code we chose an logarithmic equally spaced energy grid with 300 bins from $E=(10^{-3} - 10^{12})\,\mathrm{GeV}$. Low energy photons, called x-rays in the code, are captured on a lower energy grid with the same binning but $E_\mathrm{x-ray}=(10^{-18} - 10^{-3})\,\mathrm{GeV}$.

With that the time evolution of the number density of species $\alpha$ can be formulated by:
\begin{align}
    n_i^\alpha(t+\Delta t) = T^{\alpha \beta}_{ij}(t) n^\beta_j (t) \quad . \label{eq:numberdensity}
\end{align}
Here, $T^{\alpha \beta}_{ij}$ is the transition matrix that that describes the energy flux from species $\beta$ at energy $E_j$ to species $\alpha$ a energy $E_i$. It is composed of the normalized secondary spectra (yields) and the interaction probability. This technique is based on implementation of the \emph{matrix multiplication} method for cosmic-ray transport as described in \cite{Ray86, Ray93, Ray96} and more recently in \cite{Reimer2023}. The yields and interaction probabilities are based on well established Monte Carlo generators or analytical approximation of the individual processes (see section \ref{ssec:interactions} for details). When energy gains, due to acceleration are ignored the transition matrix is an upper triangle-matrix. 

\section{Implementation --- Framework and Design}
\label{sec:software}

When not only an electron-proton plasma but all relevant cosmic-ray nuclei should be included in a transport model this requires to track spectra of potentially several hundreds of isotopes simultaneously. Making it necessary to structure the code in such a way that the creation of transition matrices and species spectra is handled fully automatized. Therefore we decided to create a modular code, based on four physics motivated building blocks: \texttt{Species}, \texttt{Targets}, \texttt{Interactions}, and the \texttt{Simulation} class. The main code is written in python where some computation intense calculations are taken from the Fortran code \emph{CR-ENTREES} \cite{Reimer2023} and made accessible via \emph{f2py}. All tabulated data for branching ratios, interaction rates or pre-calculated yields, as well as all simulation results are stored in the hdf format, which is very efficient in disk usage, write and read times, and well supported in most modern programming languages allowing for very flexible post-processing tools. Thanks to the modular structure all parts of the code can be used and tested independent of each other. In the following we explain each of them in more detail.

\subsection{Species}
\label{ssec:species}
The \texttt{Species} class is the central structure of this software, holding all essential information of a cosmic-ray or secondary particle. This includes elementary information like a particle identifier\footnote{We use the same PDGMC identifier convention that is also used in other UHECR codes, e.g., in CRPropa \cite{CRPropa32}.}, but also arrays of the particle spectra as defined in equ.~\ref{eq:numberdensity}, and the transition matrices. In addition, information of current parent and secondary species is stored, which is useful to track interaction networks but also needed to remove species from the simulation chain (see \emph{Immediate Interaction} in section \ref{ssec:simulation}). Manipulation of spectra, adding or removing interaction, (re-) calculation of transition matrices and storing selected information for post-processing is realised with variety of different class methods. In principle all information on spectra, transition matrices, and more are accessible at any time of the simulation for all species, making testing of separate parts of the code very convenient. During a simulation run one instance of the class is created per tracked species. Species can be initialized with arbitrary source spectra, allowing for completely customizable energy dependent source spectra.

\subsection{Targets}
\label{ssec:targets}
The \texttt{Targets} class is similarly structured as the \texttt{Species} class. It mainly contains information on the energy spectrum of the target species. Currently only photon field targets are implemented (see section \ref{sec:model}) but the code allows for an easy extension to hadronic targets, e.g. gas from the interstellar medium, too. Also the photon target spectrum can be filled arbitrarily by the user, however, some convenient target field models including black body radiation or power law distributions are provided.

Non-linear models require an option to feedback produced secondary photons back into the target flux. This is done by an interpolation routine mapping the finer energy resolution of the x-ray species (20 bins per decade) to the coarser target energy grid (10 bins per decade). The \texttt{Targets} class can differentiate between external and internal target fields, where only the internal ones are subject to non-linear updates.

\subsection{Interactions}
\label{ssec:interactions}
The \texttt{Interaction} base class provides general functionality to either load pre-calculated yields and interaction rates or calculate them, when the relevant data is not available. We decided to not ship all pre-calculated yields with the software as this would lead to very large amount of data that have to be transferred together with the program code. Therefore, the new yields (not integrated over the photon target field) will be stored to speed up future simulations the first time an interaction is used for a new installation of the program. In addition to loading or calculating the raw yields and rates the \texttt{Interaction} class contains methods to perform the integration over the photon target fields, too. In the following, we describe the assumption and approximations that went into the design of each \texttt{Interaction} class. 

\paragraph{Nuclear Decay}
Decay of not only radioactive nuclei but also charged pions, kaons, and muons is included in the framework.\footnote{Neutral kaons and pions decay even at the highest Lorentz boosts faster than all relevant time scales that instead their decay products are included in the simulation.} For the nuclear decay we assume boost conservation from parent to daughter nuclei and nuclei secondary projectiles (e.g., $\alpha$-particles), leading to a diagonal transition matrix. Energy spectra for electrons/positrons are analytically calculated in the nucleus rest frame and boosted to the emission zone frame; the corresponding neutrino spectra are derived based on energy conservation. Relevant information on q-values, decay time scales and branching ratios are shipped with the code but the yields are calculated when a radioactive species is added to the simulation for the first time.

\paragraph{Photo-Meson Production}
Photo-Meson production is based on the SOPHIA $\gamma$-nucleon event generator code \cite{SOPHIA}. Yields and secondary rates for protons and neutrons are tabulated and part of the code. For heavier nuclei we apply a superposition ansatz. In this approximation, the nuclei cross section is described by the weighted sum of proton and neutron cross sections: $\sigma_{\gamma N} = f(A, Z)\sigma_{\gamma p} + f(A, A-Z)\sigma_{\gamma n}$, where $f$ is an phenomenological description of nuclear effects, $A$ is the mass and $Z$ the charge number.

\paragraph{Inverse Compton Scattering}
Inverse Compton (IC) scattering is implemented following the approximations given in \cite{Ray86}. Together with synchrotron radiation the IC process is able to produce x-rays which can be used for non-linear feedback for the photon target.

\paragraph{Electromagnetic Pair Production}
Bethe-Heitler pair production for protons is implemented following a modified approach of the description of rates and yields as given in \cite{Ray96}. For nuclei an scaling approach will be implemented where the interaction rate of nuclei scales with the square of the charge number $\tau^{-1}_\mathrm{N}=Z^2\tau^{-1}_\mathrm{p}$ and the inelasticity scales with the mass number. Pair production due to gamma-rays interacting with the target photons is included, too. For both processes yields and interaction rates are pre-calculated included with the software.

\paragraph{Synchrotron Radiation}
Synchrotron radiation is included for all charged particles. In combination with the individual treatment of muon, and charged mesons, this gives a more precise description of the emission region especially for strong magnetic fields as the energy loss time scale can be significantly smaller than the decay time scale. Interaction rates and yields are calculated for a given magnetic field strength on the fly and updated when the magnetic field changes during the simulation. Quantum correction for very strong magnetic field based on \cite{Pacho, Brainerd} are included. Electron synchrotron radiation is one of the feedback sources for the photon target field. Based on user input synchrotron self absorption on the electron population can be included as defined in \cite{Rybicki}.

\paragraph{Photo-Disintegration}
We used Talys \cite{Talys} to generate branching ratios and other relevant information for photo-disintegration. Since the number of fragmentation channels and with that the number of secondaries can grow very large, the interaction rates and yields are not shipped with the code. As for the nuclear decay, yields will be calculated when needed and stored for later use.

\subsection{Simulation}
\label{ssec:simulation}
The \texttt{Simulation} class steers the actual simulation and holds relevant global parameters such as the timestep, magnetic field strength emission region size, etc. It advances the simulation time from $t$ to $t+\Delta t$ by triggering the multiplication of each species' spectrum with their transition matrices. Furthermore, it makes sure that newly created species are added to the simulation chain and initiates the calculation of all relevant transition matrices by adding all general interactions to the new species. In a non-linear model, the simulation class will also take care of updating the target fields and re-calculate all affected interaction rates and yields. Lastly, it organizes writing spectra and further information to disk for the analysis.

\begin{figure}[htbp]
    \centering
    \includegraphics[width=\textwidth]{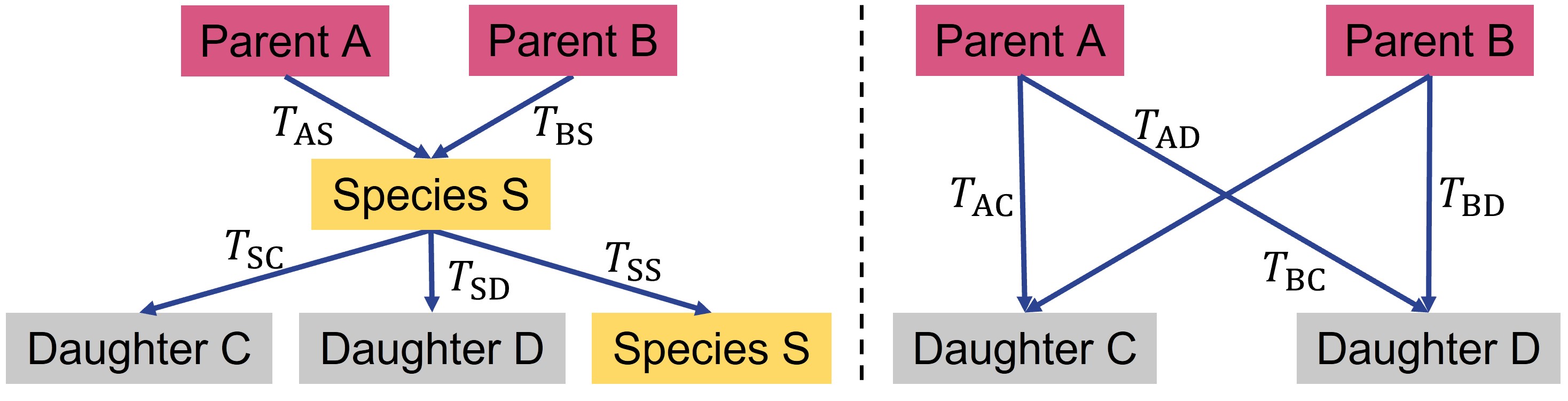}
    \caption{The left part shows the transition scheme of species $S$: It is produced by parent species $A$ and $B$ and creates secondaries species $C$ and $D$. Above an interaction probability $p>0.95$ for all energies the immediate interaction can be executed, which give the transition scheme on the right. Parent species $A$ and $B$ directly produce the daughter species $C$ and $D$.}
    \label{fig:immediateInt}
\end{figure}

\paragraph{Immediate Interaction}
To speed up the simulation it would be advantageous to remove species from the simulation chain that have very low survival probability; meaning their flux is reduced to almost zero within a single timestep, e.g., species with very short decay times. However, their secondary particles might be stable, and therefore relevant for the simulation result, and they might even be constantly produced. To remove such an unstable species $S$ consistently from the simulation chain, their parent species' ($P$) transition matrices have to be updated that they directly transfer to $S$'s daughter species $D$: $T^{PS} \rightarrow T^{PD}$ (see fig.~\ref{fig:immediateInt} for a schematic view). This procedure is handled by a method called immediate interaction and the threshold survival probability can be user defined.

\section{Testing and Examples}
\label{sec:examples}

Where available the software is validated against analytical solutions or approximations. This is especially applied to simple models where all interactions are tested separately. Since this software is an extension to the energy transport code CR-ENTREES, we make sure that the solutions agree with each other in more complex models, when only electrons and protons are considered as primary particles. The reader is referred to an upcoming paper for detailed discussion of the test cases.

\begin{figure}[htbp]
    \begin{minipage}{.49\textwidth}
        \includegraphics[width=\textwidth]{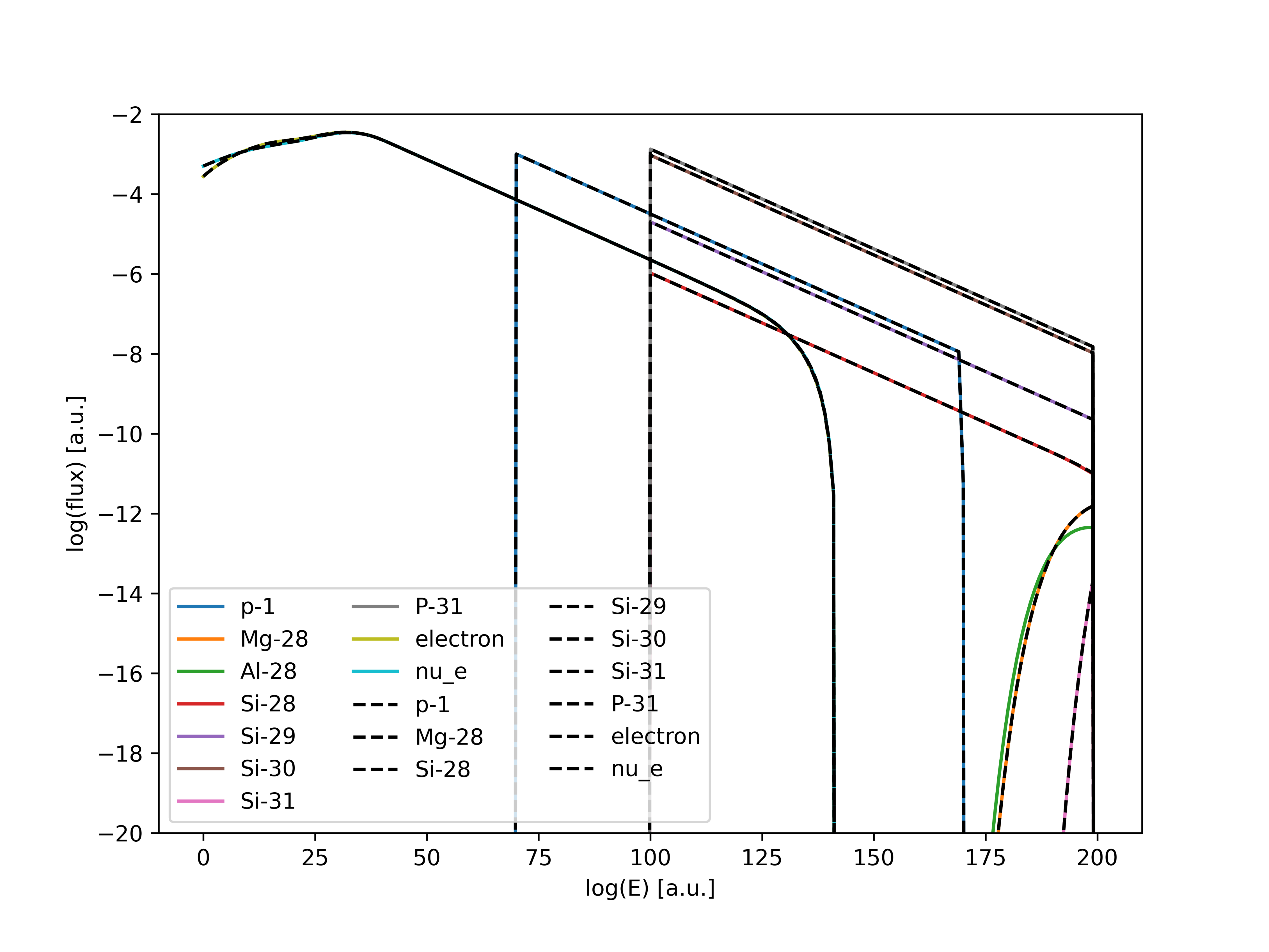}
    \end{minipage}
    \begin{minipage}{.49\textwidth}
        \includegraphics[width=\textwidth]{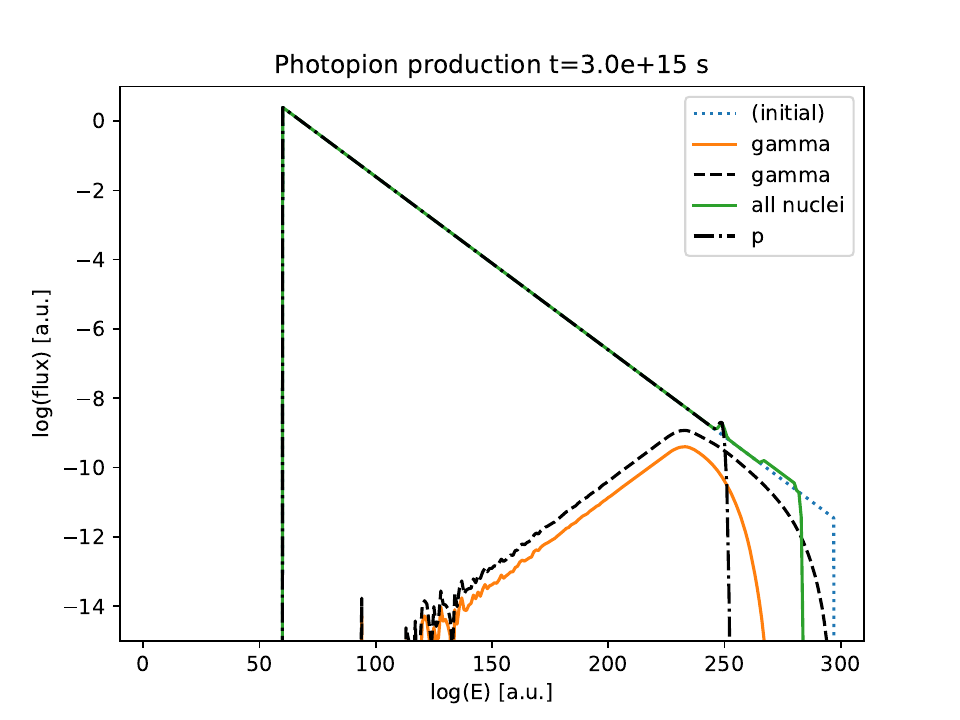}
    \end{minipage}
    \caption{Number densities of active species for the two examples. \textbf{Left plot}: Comparison of spectra produced in the nuclear decay chain of $\mathrm{Ne}^{31}$. Solid colored line correspond to a model without immediate decays, dashed black lines show results taking immediate decays into account. \textbf{Right plot}: Comparison of gamma-, and cosmic-ray spectra from photo-meson production for the injection of $\mathrm{Fe}^{56}$ (colored solid lines) and protons (black lines). Note, that nuclear was ignored in this simulation.}
    \label{fig:examples}
\end{figure}

In the following, we discuss two examples that show some of the capabilities of the new software:

\paragraph{Nuclear Decay of $\mathrm{Ne}^{31}$} The radioactive neon isotope $\mathrm{Ne}^{31}$ decays through several stages with very different decay time scales. Therefore, it makes a good candidate to test the immediate decay feature of the software. In doing so, two simulations only taking nuclear decay into account were done, one with and without the speed up. $\mathrm{Ne}^{31}$ and all decay products were modelled with a propagation time step $\Delta T = 10^{10}\,\mathrm{s}$. Figure \ref{fig:examples} shows on the left the number densities derived for the two models. No differences in stable particles are observed at time $t = 50\Delta T$.

The simulation time is decreased on the order of $\sim 10$~percent, but strongly depends on the chosen interaction probability threshold, time step and decaying element.

\paragraph{Photo-Meson Production} To emphasize the importance of models including heavy nuclei we run two simulations of photo-meson production (PMP) on a black body target field with a temperature of $T_\mathrm{BB}=27.3\,\mathrm{K}$ and a propagation time step $\Delta T = 3\times 10^{13}\,\mathrm{s}$: Once we use protons as the primary species and once we use iron. All other processes, such as photo-disintegration or nuclear decay, are excluded. Figure \ref{fig:examples} shows the spectra of the two simulation on the right. We compare the gamma-ray --- from $\pi^0$-decay --- and total cosmic-ray fluxes with each other after 100 simulation steps. On of the main reasons for the large difference in the spectra is the decreased energy per nucleon for heavy elements compared to protons at the same energy.

\section{Summary and Outlook}
\label{sec:outlook}
We presented a new simulation framework that is capable to model the fully non-linear time evolution of heavy nuclei and their secondaries, including gamma rays and neutrinos. Although the main focus of the code is an accurate description of transport and interactions in dense and time variable photon fields, as we expect them in UHECRs and neutrino sources, its modular structures allows application also in other fields of astroparticle physics. All major interaction channels with photon target fields for leptons and nuclei are included as well as synchrotron radiation, adiabatic energy losses, and particle escape. 

To deal with the large amount of species that need to be tracked for such heavy nuclei models a dynamic species management was developed that allows to add or remove species on-the-fly, which can reduce the computation time. All input and output data are stored in hdf files, making them efficiently available in many different post-processing tools. After some final optimizations in the treatment of photo disintegration the code will applied to model different astrophysical sources.

As one of the first source classes we will take a closer look at the very numerous low luminosity radio galaxies, namely Faranoff Riley (FR) 0 galaxies, which have been proposed as sources of UHECRs \cite{Merten2021}. Since they might provide a significant portion of the UHECR flux their emitted source composition is of special interest \cite{Lundquist2022}.

In addition, this software is also tailored to model time dependent neutrino emission profiles in high photon target density source more accurately than ever, as the production rates of different heavy nuclei can be directly taken into account. This will help to pre-select interesting sources for the IceCube neutrino telescope and with that reduce the amount of necessary data reduction.

%%%%%%%%%%%%%%%%%%%%%%%%%%%%%%%%%%%%%%%%%%%%%%%%%%%%%%%%%%%%%%%%%%%%%%%%%%%%%%%%%
\section*{Acknowledgements}
We thank V.~Lanzinger for her contribution to implement routines for nuclear decay and photo-disintegration.

Financial support was received from the Austrian Science Fund (FWF) under grant agreement number I 4144-N27 and the Slovenian Research Agency-ARRS (project no. N1-0111). LM and JBT acknowledge support from the DFG within the Collaborative Research Center SFB1491 "Cosmic Interacting Matters - From Source to Signal" and from the BMBF grant "05A14PC1". MB has for this project received funding from the European Union’s Horizon 2020 research and innovation program under the Marie Sklodowska-Curie grant agreement No 847476. The views and opinions expressed herein do not necessarily reflect those of the European Commission. 

%%%%%%%%%%%%%%%%%%%%%%%%%%%%%%%%%%%%%%%%%%%%%%%%%%%%%%%%%%%%%%%%%%%%%%%%%%%%%%%%%

\end{document}